\documentstyle[11pt]{article}
\title{Nonoblique Corrections in Technicolor Theories Revisited}
\author{Guo-Hong Wu\\
Department of Physics, Yale University, New Haven, CT 06520}
\begin{document}
\setlength{\baselineskip}{24pt}
\maketitle
\begin{picture}(0,0)(0,0)
\put(300,250){YCTP-P16-94}
\end{picture}
\vspace{-48pt}

\begin{abstract}

 In extended technicolor (ETC) theories, while the sideways
ETC boson exchange decreases the width
$\Gamma_b \equiv \Gamma (Z\rightarrow b \bar{b})$,
the flavor-diagonal ETC boson exchange tends to increase it, and
the ETC-corrected $R_b \equiv \Gamma_b / \Gamma_{\mbox{\footnotesize had}}$
value could agree with recent measurements.
The $\tau$ asymmetry parameter may also increase in a way
consistent with experiment.
The weak-interaction $\rho$ parameter receives a correction from
diagonal ETC exchange which is just barely acceptable by experiments.
\end{abstract}

{\small


}

  Technicolor (TC) theories with their characteristic nondecoupling
effects, often
have sizable corrections to various low energy observables that are
now under scrutiny by precise electroweak measurements.
The recent measurement of $R_b \equiv \Gamma_b /
\Gamma_{\mbox{\footnotesize had}}$
($\Gamma_{\mbox{\footnotesize had}}$ is the $Z$ hadronic width)
at the CERN $e^+e^-$ collider LEP \cite{LEP},
$R_b = 0.2202 \pm 0.0020$, already differs by more than two standard
deviations from the value $R_b^{\mbox{{\scriptsize SM}}} = 0.2157 \pm 0.0004$
predicted by the minimal standard model with the top quark mass
in the range $m_t = 163 \sim 185 \; \mbox{GeV}$ \cite{Rev}.
Previous studies [3--7]
indicate that extended technicolor (ETC) models with ETC bosons carrying
no standard model quantum numbers give negative corrections to
$R_b^{\mbox{{\scriptsize SM}}}$ and therefore have been
ruled out at high confidence-level.
Noncommuting ETC models \cite{CST} (ETC bosons carrying
weak charge) have been proposed to give a positive
correction to the standard model prediction.
More recently it has been found \cite{Hod} that an extra gauge boson
existing in certain type of models \cite{Hod2} of dynamical electroweak
 symmetry breaking can give positive corrections to both
$R_b^{\mbox{{\scriptsize SM}}}$ and the $\tau$ asymmetry parameter
$A_{\tau}$, as well as corrections to other standard model predictions,
all in agreement with recent experimental data.

  In this letter, we reconsider ETC models with ETC bosons
transforming as singlets under the standard model gauge group.
We find that while the sideways ETC boson exchange
(connecting technifermions to ordinary fermions within the same
ETC multiplet) decreases $\Gamma_b$,
the diagonal boson exchange (connecting technifermions, and ordinary
fermions, only to themselves) increases the width,
contrary to the result obtained in ref.~\cite{Kit}.
Furthermore, the two corrections are of the same order of magnitude,
and the ETC-corrected $R_b$ value could lie in
a range consistent with recent LEP measurements.
We also find that if technielectrons are much lighter than techniquarks,
the dominant contribution to the $\tau$ asymmetry parameter
$A_{\tau}$ comes from diagonal ETC boson exchange, and that
the ETC corrections may increase $A_{\tau}$ in a way consistent
with experiment.
It is worth noticing that the diagonal ETC boson considered here
 plays a similar role in electroweak radiative corrections
as the recently discussed \cite{Hod} extra gauge boson existing in
some other models \cite{Hod2} of dynamical symmetry breaking.

  The ETC models we consider have separate ETC scales for different
quark-lepton families \cite{AT2}. The third family containing the
heaviest fermions
has the lowest ETC scale and gives the largest ETC corrections to various
physical observables.  We subsequently focus on the
third family and neglect the first two.
The ``nonoblique" \cite{LPS} effects due to ETC boson exchange can be
conveniently studied in $R_b \equiv \Gamma_b/
\Gamma_{\mbox{\footnotesize had}}$,
where the ``oblique" (and QCD) corrections nearly cancel in the ratio.

   We consider a one-family TC model with technifermions belonging to
the fundamental representation of an $SU(N)_{\mbox{\scriptsize TC}}$
technicolor group and carrying the same color and electroweak quantum
numbers as their standard model counterparts.
The ordinary fermions couple to the technifermions via ETC interactions
with strengths $g_{E,L}$, $g_{E,R}^U$ and $g_{E,R}^D$ for respectively the
left-handed techniquark-quark weak doublets, the right-handed techniup
($U_R$) and $t_R$, and the right-handed technidown ($D_R$) and $b_R$
(plus couplings between the leptons and technileptons).
The ETC gauge symmetry is assumed to break down to
$SU(N)_{\mbox{\scriptsize TC}}$ below the third-family ETC scale
$m_{\mbox{\scriptsize ETC}}$, generating both sideways and diagonal ETC
bosons with masses $m_{X_S}$ and $m_{X_D}$ respectively.
The hermitian, traceless generator for the diagonal ETC boson
respects the $SU(N)_{\mbox{\scriptsize TC}}$ TC symmetry and can therefore
be normalized as $\mbox{diag}\frac{1}{\sqrt{2N(N+1)}}(1,\cdots, 1, -N)$.
The resulting effective ETC lagrangian can thus be written as
\begin{eqnarray}
{\cal L}_{\mbox{\scriptsize ETC}} & = &  {\cal L}_{\mbox{\scriptsize TC}}
        - \frac{1}{\sqrt{2}} \sum_{i=1}^{N}
       ( X^{i,\mu}_S J_{S,i,\mu}  +  X_{S,i,\mu} J_S^{i,\mu} )
        -  X_{D,\mu} J^{\mu}_D ,
\end{eqnarray}
where ${\cal L}_{\mbox{\scriptsize TC}}$ is the technicolor interaction,
 $i$ is the technicolor index,
$X^{i,\mu}_S$ and $X_{D,\mu}$ stand for the sideways
and diagonal ETC bosons respectively, and $J_{S,i,\mu}$
and $J^{\mu}_D$ denote their corresponding currents.
The sideways and diagonal ETC currents are given by
\begin{eqnarray}
J_{S,i,\mu} & = &    g_{E,L} \bar{Q}_{iL} \gamma_{\mu} \psi_L
                   + g_{E,R}^U \bar{U}_{iR} \gamma_{\mu} t_R
                   + g_{E,R}^D \bar{D}_{iR} \gamma_{\mu} b_R ,   \\
J_S^{i,\mu} & = & (J_{S,i}^{\mu})^{\dag}  ,         \nonumber     \\
J^{\mu}_D & = &
 \frac{1}{\sqrt{2N(N+1)}} g_{E,L}
(\bar{Q}_L \gamma^{\mu} Q_L - N \bar{\psi}_L \gamma^{\mu} \psi_L)  \\
 && +  \frac{1}{\sqrt{2N(N+1)}} g_{E,R}^U
(\bar{U}_R \gamma^{\mu} U_R - N \bar{t}_R \gamma^{\mu} t_R) \nonumber \\
&& +   \frac{1}{\sqrt{2N(N+1)}} g_{E,R}^D
(\bar{D}_R \gamma^{\mu} D_R - N \bar{b}_R \gamma^{\mu} b_R) , \nonumber
\end{eqnarray}
where $Q \equiv (U,D)$ is the techniquark doublet, $\psi \equiv
(t,b)$ is the quark doublet,
and summation over color (and technicolor) indices is implied.
The ETC currents in the lepton-technilepton sector are not displayed
but can be similarly written down.
The ETC corrections to the left-handed and right-handed $b$ couplings
to $Z$ ($g_L^b$ and $g_R^b$) can now be computed.
As $\delta \Gamma_b$ is more than  five times as sensitive to $\delta g_L^b$
as to $\delta g_R^b$,  only ETC corrections to $g_L^b$ will be considered.

  It has been suggested \cite{AT} that in a realistic one-family TC model,
the technileptons could be much lighter than the nearly degenerate
techniquarks in order to keep the electroweak $S$ parameter \cite{PT}
small or even negative while not to violate the experimental bound on the
$T$ parameter \cite{PT}.
The dominant contribution to the weak scale
$v= 246 \; \mbox{GeV}$ would therefore come from the techniquark sector,
\begin{equation}
v^2 = N_C f_Q^2 + \frac{1}{2} f_E^2 + \frac{1}{2} f_N^2
 \simeq N_C f_Q^2 ,
\end{equation}
where $N_C=3$ is the number of colors,  $f_Q$, $f_E$ and $f_N$ are the
Goldstone boson (GB) decay constants for the techniquark, technielectron
and technineutrino sectors respectively, and $N_Cf_Q^2 \gg (f_E^2 + f_N^2)/2$
has been used in arriving at the second expression.
We will assume this technifermion mass spectrum which makes our computations
simple and transparent, though our conclusions are not sensitive to this
assumption.

 The effective chiral lagrangian method was first used
to estimate  the sideways ETC exchange contribution to the
$Zb\bar{b}$ vertex in a one-doublet TC model by Chivukula
{\it et al.} \cite{CSS},
the same procedure will be followed here.
For the assumed technifermion mass spectrum,
the technilepton's contribution to the weak scale and
similarly to the $Zb\bar{b}$ vertex can be neglected to first approximation.

   The sideways ETC boson exchange gives rise to the following four-fermion
operators below the ETC scale $m_{X_S}$,
\begin{equation}
{\cal L}_{4f}^{S}  =  - \frac{1}{2m_{X_S}^2}
                  \sum_{i=1}^N J_S^{i,\mu} J_{S,i,\mu}
        =    - \frac{g_{E,L}^2}{2m_{X_S}^2}
   (\bar{Q}_{L} \gamma^{\mu} \psi_L)
   (\bar{\psi}_{L} \gamma_{\mu} Q_L) + \cdots ,
\end{equation}
where four-fermion operators not contributing to $g_L^b$ have been dropped,
and summation over color and technicolor indices is implied
in the second expression.
The above four-fermion operator can be Fierzed into the form
\begin{equation}
  - \frac{g_{E,L}^2}{2m_{X_S}^2}  \frac{1}{2N_C} \sum_{a=1}^{3}
(\bar{\psi}_{L} \gamma_{\mu} \tau_a \otimes 1_{3} \psi_L)
   (\bar{Q}_{L} \gamma^{\mu} \tau_a \otimes 1_{3} Q_L) + \cdots ,
\end{equation}
where $\tau_a$'s are the Pauli matrices,
and $1_3$ denotes the three by three unit matrix in color space. The
dots stand for the weak-singlet and/or color-octet pieces
that do not couple to $Z$ and thus have no contributions to the $Zb\bar{b}$
vertex.

   Below the TC chiral symmetry breaking scale, the techniquark
current is replaced by the corresponding sigma model current
\cite{Ge} of an $SU(2N_C)_L \otimes SU(2N_C)_R$ chiral symmetry
of the techniquark sector
\begin{equation}
\bar{Q}_{L} \gamma^{\mu} \tau_a \otimes 1_{3} Q_L  \rightarrow
i \frac{f_Q^2}{2} \mbox{Tr}(\Sigma^{\dag} \tau_a \otimes 1_{3} D^{\mu} \Sigma)
    \stackrel{\Sigma=1}{=}  - \frac{g}{c} Z^{\mu} N_C f_Q^2
        \frac{\delta^{3a}}{2}  + W^{\pm,\mu} \; \mbox{piece} ,
\end{equation}
where $\Sigma$ is the $2N_C$ by $2N_C$ exponentiated Goldstone boson matrix
transforming as $\Sigma \rightarrow L \Sigma R^{\dag}$ under
$SU(2N_C)_L \otimes SU(2N_C)_R$,
and $D_{\mu} \Sigma$ is the electroweak covariant derivative.
In the last expression we set the Goldstone fields to zero to
project out the $Z$ field, $g$ is the $SU(2)_L$ gauge coupling, and
$c=\cos \theta_W$ ($\theta_W$ is the Weinberg angle).

   Upon substituting Eq.~7 into Eq.~6, we can read off the correction
to $g_L^b$ from sideways ETC exchange,
\begin{eqnarray}
\delta g_L^b(\mbox{sideways}) & = & \frac{g_{E,L}^2 f_Q^2}{8m_{X_S}^2}.
\end{eqnarray}
Since the standard model tree level value
$g_L^b = - \frac{1}{2} + \frac{1}{3} s^2$ ($s^2 \equiv \sin^2 \theta_W$)
 is negative,
the sideways ETC exchange decreases $\Gamma_b$ relative to the standard
model prediction.
It is worth emphasizing \cite{App} that the result of Eq.~8 is
of complete
generality rather than a low energy effective lagrangian approximation,
and that it is directly related to whatever (TC) dynamics contributes the
$Z$ self energy (the weak scale). The chiral lagrangian used above
 is nothing more than one convenient way of providing this connection.
The same applies to Eq.~10 below.

Below the ETC scale $m_{X_D}$, the diagonal boson exchange gives rise to
the following four-fermion operators,
\begin{equation}
{\cal L}_{4f}^{D}  =  - \frac{1}{2m_{X_D}^2}
                  J_{D,\mu} J_D^{\mu}
        =  \frac{1}{4m_{X_D}^2} \frac{1}{N+1} g_{E,L}
      (g_{E,R}^U - g_{E,R}^D)
(\bar{Q}_R \tau_3 \gamma^{\mu} Q_R)(\bar{\psi}_L \gamma_{\mu} \psi_L)
  + \cdots ,
\end{equation}
where color and technicolor summation is implied, and
we have retained only the dominant techniquark contribution to
the $Zb\bar{b}$ vertex.
   Below the TC chiral symmetry breaking scale,
the right-handed techniquark current is replaced by the corresponding
sigma model current
\begin{equation}
\bar{Q}_R \tau_3 \otimes 1_3 \gamma^{\mu} Q_R  \rightarrow
i \frac{f_Q^2}{2} \mbox{Tr}(\Sigma \tau_3 \otimes 1_{3} (D^{\mu}\Sigma)^{\dag})
  \\
   \stackrel{\Sigma=1}{=}   \frac{g}{c} Z^{\mu} \frac{N_C f_Q^2}{2} .
\end{equation}
Substituting Eq.~10 into Eq.~9, the correction to
$g_L^b$ from diagonal ETC boson exchange can be read off,
\begin{eqnarray}
\delta g_L^b(\mbox{diagonal}) & \simeq & - \frac{f_Q^2}{8m_{X_D}^2}
   \frac{N_C}{N+1} g_{E,L} (g_{E,R}^U - g_{E,R}^D) .
\end{eqnarray}

  The result of Eq.~11 differs by a minus sign from the loop estimate of
ref.~\cite{Kit}, and it arises from the opposite-sign couplings
of the diagonal boson to techniquarks and quarks
(we have checked the sign by doing the loop calculation).
It is noted that the masses of $t$ and $b$ are related to
the techniquark condensates by
 $m_t=\frac{g_{E,L} g_{E,R}^U <\bar{U}U>}{2m_{X_S}^2}$ and
 $m_b=\frac{g_{E,L} g_{E,R}^D <\bar{D}D>}{2m_{X_S}^2}$
respectively, and that $m_t \gg m_b$ can
be explained only if $g_{E,R}^U > g_{E,R}^D$
(we take the ETC couplings $g_E$'s to be positive for simplicity,
though they are only required to be of the same sign).
 The diagonal ETC exchange therefore gives a negative
correction to $g_L^b$ and increases the width $\Gamma_b$,
contrary to the sideways ETC exchange.
Note that the size of the diagonal ETC exchange correction becomes smaller
with increasing number of technicolor.

  Summing up the sideways and diagonal ETC exchange contributions gives
\begin{eqnarray}
\delta g^b_{L,\mbox{\scriptsize ETC}} & \simeq & - \frac{f_Q^2}{8}
 [ \frac{g_{E,L} (g_{E,R}^U - g_{E,R}^D)}{m_{X_D}^2} \frac{N_C}{N+1}
  - \frac{g_{E,L}^2}{m_{X_S}^2} ]    \\
 & \stackrel{N=2}{\simeq} &  - \frac{v^2}{24 m_{X_S}^2}
 [ \frac{m_{X_S}^2}{m_{X_D}^2} g_{E,L} (g_{E,R}^U - g_{E,R}^D)
  - g_{E,L}^2] .        \nonumber
\end{eqnarray}
It is seen from the above expression that the two contributions are of
comparable magnitude,
and that for large $N$ the sideways ETC exchange contribution dominates and
the width $\Gamma_b$ decreases.
Both the electroweak $S$ parameter
 and current LEP result of $R_b$ favor small values of $N$.
We have therefore taken $N=2$ in arriving at the second expression and
will assume this value throughout this paper.
The size of the ETC couplings $g_E$'s and
the ratio of the masses of the sideways and diagonal ETC bosons
are ETC-model-dependent.
It is therefore possible for ETC exchange to give a positive correction
to $R_b$.

 So far the focus has been on the ``high energy" ETC
contributions to the $Zb\bar{b}$ vertex from ``integrating out"
technifermions, there are also ``low energy" contributions
coming from the pseudo-Goldstone-bosons (PGB's).
For a one-family QCD-like TC model and including only color contributions to
the masses of color-octet PGB's, these effects have been found to
decrease $R_b$ by a few percent \cite{Ch}.
In realistic ETC models, masses of the PGB's could be significantly
enhanced by whatever dynamics enhances the technifermion condensates,
as in ``walking" TC theories \cite{walking} and in strong-ETC models
\cite{sETC}, and the PGB correction to $R_b$ may significantly be reduced.

  We now use Eq.~12 for an estimate of the TC correction to $R_b$,
taking $N=2$ and
assuming the dominant contributions to the $Zb\bar{b}$ vertex come
from ETC exchange.
Denoting the generic ETC couplings by $g_E$ and ETC
 boson masses by $m_{\mbox{\scriptsize ETC}}$,
we have $\delta g_{L,\mbox{\scriptsize ETC}}^b \sim - \frac{v^2}{24}
\frac{g_E^2}{m_{\mbox{\scriptsize ETC}}^2}$ from diagonal ETC exchange.
The diagonal ETC correction to $R_b$ is then
\begin{equation}
\frac{\delta R_b}{R_b}
     \simeq  (1- R_b) \frac{2 g_L^b \delta g_L^b}
                     {{g_L^b}^2 + {g_R^b}^2}
  \sim  0.9 \% \times \frac{g_E^2}
{(m_{\mbox{\scriptsize ETC}}/{\mbox{TeV}})^2} ,
\end{equation}
where the value $s^2=0.232$ has been used.
In order for the diagonal exchange alone to result in an effect as large as
seen at LEP, it is necessary that
$g_E^2/m_{\mbox{\scriptsize ETC}}^2 \sim (2 \pm 1)/\mbox{TeV}^2$.
In strong-ETC models \cite{sETC} where the
large $t$ mass is generated with a near-critical ETC coupling
($g_E^2/ 4 \pi^2 \simeq 1$), LEP result of $R_b$ would require
$m_{\mbox{\scriptsize ETC}} \sim 3 \ \mbox{--} \ 6 \; \mbox{TeV}$.
  In this case, the strong ETC dynamics contributes an essential part of the
$t$ mass, whereas to a large extent the $Zb\bar{b}$ vertex is only sensitive
to the TC dynamics, and unlike QCD-scaled-up TC models \cite{CSS}
there is no direct connection between $R_b$ and $m_t$ \cite{Ev}.

   The sensitivity of the $\tau$ asymmetry
$A_{\tau} \equiv \frac{{g_L^{\tau}}^2 - {g_R^{\tau}}^2}
{{g_L^{\tau}}^2 + {g_R^{\tau}}^2}$
($g_L^{\tau}= -\frac{1}{2} + s^2$ and $g_R^{\tau}= s^2$ are the
left-handed and right-handed $\tau$ couplings to $Z$)
to new physics can be appreciated
by noting that
\begin{equation}
\frac{\delta A_{\tau}}{A_{\tau}}  =
\frac{4{g_L^{\tau}}^2{g_R^{\tau}}^2}{{g_L^{\tau}}^4 - {g_R^{\tau}}^4}
           ( \frac{\delta g_L^{\tau}}{g_L^{\tau}}
           - \frac{\delta g_R^{\tau}}{g_R^{\tau}} )
         =  -25.5 \; \delta g_L^{\tau} - 29.5 \; \delta g_R^{\tau}  ,
\end{equation}
where $s^2=0.232$ has been used and
 the large coefficients are due to the $1/(1-4s^2)$ enhancement factor.
 Like the $Zb\bar{b}$  vertex,  $Z\tau \bar{\tau}$ vertex receives
corrections from both sideways and diagonal ETC exchange
(for the assumed technifermion mass spectrum \cite{AT} where the
resulting PGB's after TC chiral symmetry breaking contain no
techni-leptoquarks,
the PGB corrections to the $Z\tau \bar{\tau}$ vertex will be suppressed
by the square of the small $\tau$ mass and can thus be neglected).
The sideways ETC exchange involves technielectrons whereas for
diagonal ETC exchange the dominant contribution comes from techniquarks.
A similar calculation to that of the $Zb\bar{b}$ vertex gives
\begin{eqnarray}
\delta g^{\tau}_{L,\mbox{\scriptsize ETC}} & \simeq & - \frac{f_Q^2}{8}
 [ \frac{g_{E,L}^{\tau} (g_{E,R}^U - g_{E,R}^D)}{m_{X_D}^2} \frac{N_C}{N+1}
  - \frac{{g_{E,L}^{\tau}}^2}{m_{X_S}^2} \frac{f_E^2}{f_Q^2}]
   \\
\delta g^{\tau}_{R,\mbox{\scriptsize ETC}} & \simeq & - \frac{f_Q^2}{8}
 [ \frac{g_{E,R}^{\tau} (g_{E,R}^U - g_{E,R}^D)}{m_{X_D}^2} \frac{N_C}{N+1}
  + \frac{{g_{E,R}^{\tau}}^2}{m_{X_S}^2} \frac{f_E^2}{f_Q^2}]
\end{eqnarray}
where  $g_{E,L}^{\tau}$ and $g_{E,R}^{\tau}$ are the ETC couplings for
 $\tau_L$ and $\tau_R$ respectively.

  It is seen from Eqs.~14 -- 16 that ETC exchange in general increases
$A_{\tau}$ relative to the standard model prediction.
For our assumed technifermion mass spectrum
where $f_E^2 \ll f_Q^2$ \cite{AT},
the diagonal ETC exchange dominates over the sideways exchange and a simple
estimate of $\delta A_{\tau}/A_{\tau}$ can be made by
assuming that the ETC couplings are of comparable magnitude (the fermion mass
spectrum could partly arise from the hierarchy in the
technifermion condensates \cite{AT2}). We then get (taking $N=2$ and
$g_E^2/m_{\mbox{\scriptsize ETC}}^2 \sim (2 \pm 1)/\mbox{TeV}^2$),
\begin{equation}
\delta g_{L,\mbox{\scriptsize ETC}}^{\tau} \sim \delta
g_{L,\mbox{\scriptsize ETC}}^b \sim
- (5.0 \pm 2.5) \times 10^{-3}, \;\;\;\;\;\;\;\;\;\;
\delta g_{R,\mbox{\scriptsize ETC}}^{\tau} \sim \delta
 g_{L,\mbox{\scriptsize ETC}}^{\tau} .
\end{equation}
And the ETC exchange correction to $A_{\tau}$ is then
$\delta A_{\tau}/A_{\tau} \sim 0.28 \pm 0.14$.
Note that the ETC correction to $A_{\tau}$ could be significantly
reduced if $\tau$ couples to the technifermion sector
at a higher ETC scale than the $t$ quark.
Experimentally, $\delta A_{\tau}/A_{\tau}$ can be extracted \cite{Hod}
from lepton asymmetry measurements at LEP \cite{tau}:
assume $e$, $\mu$ universality, and average
$A_{\mbox{\scriptsize FB}}^{0 \; \tau}/A_{\mbox{\scriptsize FB}}^{0 \; e,\mu}$
from lepton
forward-backward asymmetries and
$3P_{\tau}/4P^{\mbox{\scriptsize FB}}_{\tau}$ from
$\tau$ polarization (both are equal to $A_{\tau}/A_e$
and are insensitive to oblique corrections). Comparing the average
to unity gives $\delta A_{\tau}/A_{\tau}(\mbox{exp}) = 0.14 \pm 0.10$.
Thus future experimental improvements on the precise lepton asymmetry
measurements will start constraining ETC models which could give large
positive corrections to $A_{\tau}$.

  On dimensional grounds, diagonal ETC exchange could contribute significantly
to the precisely constrained $\rho$ parameter just like to $Zb\bar{b}$ coupling
\cite{Ack}.
The dominant contribution comes from the four techniquark operator \cite{ABCH},
\begin{eqnarray}
{\cal L}_{4f}^{\Delta \rho}  & = &
   - \frac{1}{16N(N+1)} \frac{(g_{E,R}^U - g_{E,R}^D)^2}{m_{X_D}^2}
  (\bar{Q}_R \tau_3 \gamma^{\mu} Q_R) (\bar{Q}_R \tau_3 \gamma_{\mu} Q_R),
\end{eqnarray}
where the prefactor comes from the diagonal ETC coupling strengths.
The correction to the $\rho$ parameter can then be read off
after substituting Eq. 10 into the above expression,
\begin{equation}
\Delta \rho_{\mbox{\scriptsize ETC}}  \simeq
   \frac{v^2}{8N(N+1)} \frac{(g_{E,R}^U - g_{E,R}^D)^2}{m_{X_D}^2}
  \simeq 0.13 \% \times  \frac{(g_{E,R}^U - g_{E,R}^D)^2}
  {(m_{X_D}/{\mbox{TeV}})^2}.
\end{equation}
For $(g_{E,R}^U - g_{E,R}^D)^2/m_{X_D}^2 \sim
g_E^2/m_{\mbox{\scriptsize ETC}}^2 \sim (2 \pm 1)/\mbox{TeV}^2$,
this gives $\Delta \rho_{\mbox{\scriptsize ETC}} \sim (0.26 \pm 0.13) \%$.
Combined with the technicolor-sector contributions estimated in the one family
TC model \cite{AT}, this is barely consistent with recent global fits to
data \cite{rhoexp}. On the other hand, the ETC correction to
the $S$ parameter is found to be negligible compared to the TC contributions.

  In conclusion, diagonal ETC exchange with
$g_E^2/m_{\mbox{\scriptsize ETC}}^2 \sim (2 \pm 1)/\mbox{TeV}^2$ could
explain the LEP $R_b$ measurement if it is the dominant contribution
in conventional ETC models, this in turn gives a positive correction
to the $\rho$ parameter which is just barely acceptable by experiments.
The $\tau$ asymmetry parameter $A_{\tau}$ could receive a large and positive
correction from diagonal ETC exchange if the $\tau$ couples at the same ETC
scale as the top quark, and further improvements on precision lepton asymmetry
measurements at LEP should provide very interesting information on
the lepton sector in ETC theories.

\section*{Acknowledgments}

  I am grateful to Prof. T. Appelquist for suggesting this
topic, reading the manuscript, and for his valuable discussions and
suggestions.  I would like to thank W.~Marciano for interesting and
illuminating conversations and for bringing ref.~\cite{tau} to my attention.
Helpful discussions with N.~Evans and S.~Selipsky are deeply appreciated.
I would also like to thank B. Holdom for clarifying comments.
This work was supported by DOE Grant No. DE-AC02ERU3075.

\end{document}